\documentclass{elsart3}

\usepackage{amssymb}
\usepackage{graphicx}

\begin{document}

\begin{frontmatter}



\title{Low-lying magnon excitations in integer-spin ladders and tubes}


\author{Masahiro Sato}

\address{Department of Physics, Tokyo Institute of Technology, 
Oh-okayama, Tokyo 152-8550, Japan}

\begin{abstract}
We consider low-energy excitation structures of $N$-leg integer-spin 
ladder and tube systems with an antiferromagnetic (AF) intrachain 
coupling and a uniform external field.
The tube means the ladder with the periodic
boundary condition along the interchain (rung) direction. 
Odd-leg AF-rung tubes have the frustration. 
In order to analyze all systems including frustrated 
tubes, we apply a field-theoretical method 
based on the nonlinear sigma model. 
We mainly focus on the systems without any external fields.
In this case, it is shown that the lowest bands 
of frustrated tubes always consist of six-fold degenerate magnon 
excitations, while those of all other systems are triply degenerate.
This result implies that 
the ground states of frustrated tubes (all non-frustrated systems) become 
a two (one)-component Tomonaga-Luttinger liquid, 
when a sufficiently strong uniform field is applied.
\end{abstract}

\end{frontmatter}

\section{Introduction}
\label{sec1}
\begin{figure}[t]
\begin{center}\leavevmode
\includegraphics[width=0.5\linewidth]
{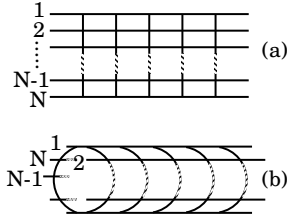}
\end{center}
\caption{$N$-leg spin ladder (a) and tube (b).}
\label{fig1}
\end{figure}

Quasi-one-dimensional (1D) spin systems 
have attracted much interest theoretically and experimentally for a long
time. Recently, spin ladders~\cite{Dagg}, 
finite-number coupled spin chains [see Fig.~\ref{fig1}(a)], 
have been studied intensively. 
In spin ladders with antiferromagnetic (AF) intrachain coupling, 
it is widely believed that odd-leg and half-integer-spin 
ladders have a gapless excitation, while other ladders possess a spin
gap. A main theoretical support of this prediction is given by 
the nonlinear sigma model (NLSM) method~\cite{Hal,Sier,Dell}. 
Moreover, it is also verified by several experiments~\cite{Hiroi} 
and numerical methods~\cite{Dagg,White,Fri}. 
As a natural extension of spin ladders, one can imagine 
spin tubes, 
which mean spin ladders with the periodic boundary condition 
for the interchain (rung) direction.
In contrast to spin ladders, 
the simple spin tubes as in Fig.~\ref{fig1}(b) have not been synthesized yet. 
However, quite recently, a few spin-tube-like
materials~\cite{Nojiri,Mill} have been reported. 
It will accelerate the theoretical study of spin tubes. 
It is known that the above even-odd property of ladders 
breaks down in tubes: the odd-leg spin-$\frac{1}{2}$ tube with AF 
intrachain and rung couplings has doubly degenerate ground states 
with a spin gap~\cite{Taka}. 

Among these spin ladders and tubes, the 2-leg ladders have been
studied most intensively. 
On the other hand, 
our understanding of ladders and tubes with higher 
chain numbers and spin magnitudes have not been progressed well, except
for the above even-odd property.
In this context, we consider the low-energy physics of
simple spin ladders and tubes, models of Fig.~\ref{fig1}, in this paper.  
Odd-leg AF-rung tubes have the frustration along the rung. 
For such frustrated tubes, the standard NLSM
scheme, which assumes some short-range order to occur for the rung
direction, is not applicable.   
In order to treat all systems containing frustrated tubes, we
construct an alternative NLSM method, a simple extension of 
S\'en\'echal's method~\cite{Sene}, which is a weak rung-coupling approach.    
As well known, a topological term appears in the NLSM 
for the half-integer-spin systems. 
Since it makes the analysis fairly complicated,
we consider only integer-spin systems.

\section{Model and method}
The spin Hamiltonian, which we investigate, is
\begin{eqnarray}
\label{eq1}
{\mathcal H} & =&   
J \sum_{n=1}^N \sum_j {\bf S}_{n,j} \cdot {\bf S}_{n,j+1}
\nonumber\\
&&+ J_\perp \sum_{n=1}^M \sum_j {\bf S}_{n,j} \cdot {\bf S}_{n+1,j}
-h\sum_{n=1}^N\sum_j S_{n,j}^z,
\end{eqnarray}
where ${\bf S}_{n,j}$ is the integer-spin-$s$ operator on 
site $j$ of the $n$th chain, 
and $M=N-1$ ($M=N$) for $N$-leg spin ladders 
(tubes: ${\bf S}_{N+1,j}={\bf S}_{1,j}$). The intrachain coupling
$J$ is positive, i.e., AF. The final term represents the uniform Zeeman
energy. 

Let us explain the strategy dealing with the model (\ref{eq1}) below.
We start from the integer-spin-$s$ AF chain: 
$H_n=J\sum_j{\bf S}_{n,j} \cdot {\bf S}_{n,j+1}$. Its
low-energy physics can be described by an O(3) NLSM.
The partition function $Z$ and Euclidean action $A$ are 
\begin{eqnarray}
Z &=& \int D{\bf n}D\lambda \,\,\exp(-A[{\bf n};\lambda]),\\
A &=& \int d{\bf x} \,\,\frac{-1}{2gc}\,{\bf n}\cdot
(\partial_\tau^2+c^2\partial_x^2){\bf n}-i\lambda({\bf n}^2-1), 
\end{eqnarray}
where $d{\bf x}=d\tau dx$ ($\tau$: imaginary time, $x$: spatial coordinate), 
${\bf n}({\bf x})$ is a vector field with three components,
$g=2/s$ is the coupling constant, $c=2sJa_0$ ($a_0$: lattice constant) is
the spin-wave velocity, and $\lambda({\bf x})$ is the Lagrange multiplier for 
the constraint ${\bf n}^2=1$. The spin operator is associated with the
field ${\bf n}$ through 
\begin{eqnarray}
{\bf S}_{n,j}&\approx &\frac{i}{4J}({\bf n}\times \dot{\bf n})+ (-1)^js{\bf n},
\end{eqnarray}
where $\dot{\bf n}=\partial_\tau {\bf n}$.
A simple estimation method of the lowest
excitation energy of $H_n$, i.e., the Haldane gap, is the saddle-point
approximation (SPA) for $\lambda$. It replaces the field $\lambda$ with
a constant $\lambda_{sp}$, which is determined by the saddle-point
equation (SPE), 
\begin{eqnarray}
\label{spe}
\partial A[\lambda_{sp}]/\partial\lambda_{sp}=0,  
\end{eqnarray}
where $A[\lambda]$ is defined as 
$e^{-A[\lambda]}=\int D{\bf n}\,e^{-A[{\bf n};\lambda]}$. 
Employing the Fourier transformation, ${\bf n}({\bf x})\propto
\sum_{\omega_n,k}e^{i(kx-\omega_n\tau)}\tilde{\bf n}(\omega_n,k)$, one
can easily calculate $A[\lambda_{sp}]$, and then Eq.~(\ref{spe}) becomes 
\begin{eqnarray}
\label{spe-0K}
\frac{3g}{2\pi}\ln\Big[\Lambda\xi+\sqrt{1+\Lambda^2\xi^2}\Big]&=&1,
\end{eqnarray}
at zero temperature~\cite{Erco}. 
Here, $\Lambda$ is the ultraviolet cut off of the 
wave number $k$, and $\xi^{-2}=-2ig\lambda_{sp}/c$. 
Through this SPA, the NLSM is mapped to a field theory of
triply-degenerate bosons $n^{x,y,z}$ with the dispersion
$\epsilon(k)=c\sqrt{k^2+\xi^{-2}}$. 
These bosons may be interpreted as the lowest spin-1 magnons. 
Equation~(\ref{spe-0K}) determines the Haldane gap $\Delta(s)=c\xi^{-1}$
as
\begin{eqnarray}
\label{gap}
\Delta(s) &=& c\Lambda/\sinh(s\pi/3).
\end{eqnarray}
Since the accurate values of $\Delta(s)$ can be provided 
by numerical methods~\cite{Todo},
Eq.~(\ref{gap}) inversely fixes the cut off $\Lambda$.
Therefore, NLSM plus SPA is a one-parameter-fitting theory~\cite{Erco}.

We next take into account the rung-coupling term perturbatively. 
Namely, in the NLSM framework, we approximate it as 
\begin{eqnarray}
\label{rung}
\sum_{n=1}^M\sum_j{\bf S}_{n,j}\cdot{\bf S}_{n+1,j} 
\to 
\frac{s^2}{a_0}\sum_{n=1}^M\int dx \,\,{\bf n}_n\cdot{\bf n}_{n+1},
\end{eqnarray}
where the subscript $_n$ means the chain index. 
This prescription was first proposed by S\'en\'echal 
in the analysis of 2-leg ladders. It is clear that Eq.~(\ref{rung}) is
efficient even for frustrated tubes.  
From Eq.~(\ref{rung}), it is also found that the total action of the
model (\ref{eq1}) is a bilinear form of ${\bf n}_n$ like the
single-chain case.
On the other hand, we have to treat $N$ constraints, 
${\bf n}_n^2=1$, differently from the single chain. 
Since the exact treatment of $N$ constraints is considerably difficult, 
we replace them to an averaged one, 
\begin{eqnarray}
\label{ave-const}
\sum_{n=1}^N{\bf n}_n^2 &=& N.  
\end{eqnarray}
Relying on Eqs.~(\ref{rung}) and (\ref{ave-const}), we study the
model (\ref{eq1}) in the following sections.

One should notice the following aspects. 
(i) The perturbative treatment (\ref{rung}) is valid only for the weak
rung-coupling region. 
(ii) The averaging of constraints as in Eq.~(\ref{ave-const}) would be a
less accurate approximation for larger-$N$. 

\section{Cases without external fields}
\begin{figure}[t]
\begin{center}\leavevmode
\includegraphics[width=0.4\linewidth]{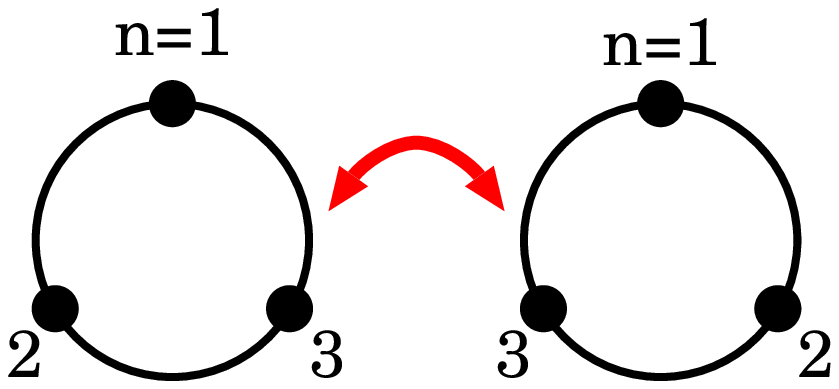}
\end{center}
\caption{Cross section and reflecting operation in $3$-leg spin tube.}
\label{fig2}
\end{figure}

\begin{figure}[t]
\begin{center}\leavevmode
\includegraphics[width=0.7\linewidth]{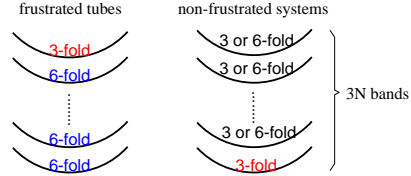}
\end{center}
\caption{Magnon band structures of $N$-leg systems. ``$\gamma$-fold''
 means the band has a $\gamma$-fold degeneracy. Six-fold degenerate
 bands do not appear in ladders.}
\label{fig3}
\end{figure}

This section discusses the model (\ref{eq1}) with $h=0$.
As mentioned already, the effective action of (\ref{eq1}) is a 
bosonic bilinear form under the approximations (\ref{rung}) and 
(\ref{ave-const}). Therefore, even for ladders or tubes, 
we can utilize an SPA scheme in a similar way of the single-chain case. 
If we represent the Lagrange multiplier for 
Eq.~(\ref{ave-const}) by the same symbol $\lambda$, its SPE becomes, 
at zero temperature,
\begin{eqnarray}
\label{spe-Nleg}
\frac{3g}{2\pi}\sum_l\ln\Big[ \Lambda\xi_l+\sqrt{1+\Lambda^2\xi_l^2} \Big] &=&   N,
\end{eqnarray}
where $\xi_l^{-2}=\xi^{-2}+2J_\perp\cos k_l/(Ja_0^2)$ and
$k_l=\frac{\pi l}{N+1}$ ($l=1,\dots,N$) [$k_l=\frac{2\pi l}{N}$
($-[\frac{N}{2}]<l\leq [\frac{N}{2}]$)]  for ladders [tubes]. (The symbol
$[\alpha]$ stands for the maximum integer $\beta$ satisfying
$\beta\leq \alpha$.) In tubes, $k_l$ means the wave number
of the rung direction.
Here, we used the same cut off $\Lambda$ as that of the single chain. 
This SPA leads to $3N$ magnon bands with dispersions 
$\epsilon_l(k)=c\sqrt{k^2+\xi_l^{-2}}$. 
The band splitting $\epsilon(k)\to \epsilon_l(k)$ is, of course, 
yielded from the hybridization among magnons due to the rung coupling.
Each band $\epsilon_l(k)$ has
a triple degeneracy, which corresponds to a spin-1 magnon triplet
($S^z=1,0$ and $-1$). Bands of {\em tubes} possesses
{\em six-fold} degeneracy $\epsilon_l(k)=\epsilon_{-l}(k)$ except for 
$\epsilon_0(k)$ and $\epsilon_{N/2}(k)$. This extra degeneracy
originates from the symmetry under the reflection about the plane
including the central axis of tubes (see Fig.~\ref{fig2}).
Paying attention to these properties of $\epsilon_l(k)$, 
we can draw the magnon band structure as in Fig.~\ref{fig3}.
It is remarkable that {\em lowest bands of frustrated tubes have
a six-fold degeneracy}, whereas all other systems have a standard triply
degenerate lowest bands.
 
\begin{figure}[t]
\begin{center}\leavevmode
\hspace{0.2cm}
\includegraphics[width=0.66\linewidth]{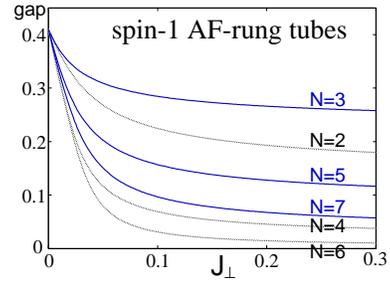}
\end{center}
\caption{Lowest excitation gaps in $N$-leg spin-1 AF-rung tubes with $J=1$. 
In order to fix $\Lambda$ in Eq.~(\ref{spe-Nleg}), we used the quantum
 Monte Carlo (QMC) data, $\Delta(1)=0.410\times J$~\cite{Todo}. The ``2-leg'' tube
 means the 2-leg ladder.}
\label{fig4}
\end{figure}
If we determine $\xi$ from Eq.~(\ref{spe-Nleg}), we can discuss 
the band structure in more detail. Figure~\ref{fig4} represents the
rung-coupling dependence of lowest excitation gaps in $N$-leg spin-$1$
AF-rung tubes.
It shows that gaps decrease with increasing $J_\perp$ and $N$, and
gap reductions in frustrated (odd-leg) tubes are slower than those in
non-frustrated tubes. The latter must reflect the fact that 
the growth of the AF short-range order is disturbed in frustrated tubes.
Although all gaps monotonically decrease with increasing $J_\perp$ within the
present scheme, one should recall that the scheme is valid especially for the
weak rung-coupling region. Actually, it is known and easily
expected that the gap growth of the 2-leg spin-1 AF-rung ladder turns
from a decrease to an increase 
as $J_\perp$ exceeds $O(J)$~\cite{Matu}.

\begin{figure}[t]
\begin{center}\leavevmode
\includegraphics[width=0.8\linewidth]{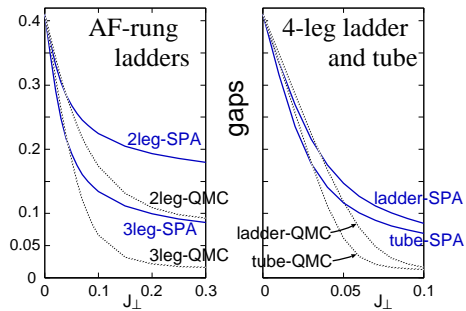}
\end{center}
\caption{Gaps of SPA and QMC methods in spin-1 AF-rung ladders or tubes 
with $J=1$. The QMC data is provided by Munehisa Matsumoto.}
\label{fig5}
\end{figure}

Finally, let us compare our results with the ones of the QMC
method. From Fig.~\ref{fig5}, it is expected that our results are
semi-quantitatively correct within $|J_\perp|<0.05\times J$, and 
are qualitatively reliable even for the weak rung-coupling case 
of $|J_\perp|>0.05\times J$.  
The main origin of the deviation between two data would lie in 
approximations~(\ref{rung}) and~(\ref{ave-const}), and the SPA. 

\section{External-field effects}
We consider the effect of the uniform Zeeman term briefly.
To do so, we should note the following general features.
(i) All excitations with $\epsilon_l(k)$ are regarded as spin-1 magnons.
(ii) The uniform field $h$ in Eq.~(\ref{eq1}) induces the Zeeman
splitting of spin-1 magnons, and the magnon band with $S^z=+1$ goes
down linearly with $h$. (iii) In general, when a magnon condensation
occurs in a 1D spin system conserving the $z$ component of the total
spin, the ground state becomes one-component
Tomonaga-Luttinger liquid (TLL)~\cite{Aff,Kon}. 

Taking into account (i)-(iii) and
the band structure in the previous section, one can expect that a sufficiently
strong field $h$ induces a two (one)-component TLL for frustrated tubes
(non-frustrated ladders and tubes).

\section{Summary}
We investigated low-lying magnon excitations of integer-spin ladders and
tubes~(\ref{eq1}) in the weak rung-coupling region. 
It was found that lowest excitations of frustrated
tubes consist of six-fold degenerate magnon bands, while those of all other
systems are standard triply degenerate ones. In addition, we
can also expect that a two (one)-component TLL emerges in frustrated
tubes (non-frustrated systems), as the field $h$ increases enough.
This expectation, however, is based on the analysis, in which the
field $h$ does not reach the lower critical value. We will revisit the
effects of the Zeeman term via other approaches~\cite{Aff,Kon,MS-nonabel}, 
elsewhere.
Furthermore, we will discuss more detail of our method here and effects
of other external fields elsewhere.

\section*{Aknowledgements}
The author thanks Masaki Oshikawa for fruitful discussions, and also does
Munehisa Matsumoto for giving the QMC data (Fig.~\ref{fig5}).





\begin{thebibliography}{00}

\bibitem{Dagg}E. Dagotto and T. M. Rice, Science {\bf 271}, 618 (1996).

\bibitem{Hal}F.D.M. Haldane, Phys. Rev. Lett. {\bf 50}, 1153
	(1983).
\bibitem{Sier}G. Sierra, J. Phys. A {\bf 29}, 3299 (1996).
\bibitem{Dell}S. Dell'Aringa, {\em et al}, Phys. Rev. Lett. {\bf 78},
	2457 (1997).

\bibitem{Hiroi}Z. Hiroi, {\em et al}, J. Solid State Chem. {\bf 95} 230
	(1991); M. Azuma, {\em et al}, Phys. Rev. Lett. {\bf 73}, 3463
	(1994).


\bibitem{White}S. R. White, {\em et al}, Phys. Rev. Lett. {\bf 73}, 886
	(1994). 
\bibitem{Fri}B. Frischmuth, {\em et al}, Phys. Rev. B {\bf 55}, R3340
	(1997).

\bibitem{Nojiri}J. Schnack, {\em et al}, Phys. Rev. B {\bf 70}, 174420
	(2004).
\bibitem{Mill}P. Millet, {\em et al}, J. Solid State Chem. {\bf 147},
	676 (1999).


\bibitem{Taka}K. Kawano and M. Takahashi, J. Phys. Soc. Jpn. {\bf 66}
	4001 (1997).

\bibitem{Sene}D. S\'en\'echal, Phys. Rev. B {\bf 52}, 15139 (1995).










\bibitem{Erco}E. Ercolessi, {\em et al}, Phys. Rev. B {\bf 62}, 14860
	(2000); Europhys. Lett. {\bf 52}, 434 (2000). 
\bibitem{Todo}S. Todo and K. Kato, Phys. Rev. Lett. {\bf 87},
	047203 (2001).

\bibitem{Matu}S. Todo, {\em et al}, Phys. Rev. B {\bf 64}, 224412
	(2001).

\bibitem{Aff}I. Affleck, Phys. Rev. B {\bf 43}, 3215 (1991);
	E. S. S\o rensen and I. Affleck, Phys. Rev. Lett. {\bf 71}, 1633
	(1993). 
\bibitem{Kon}R. M. Konik and P. Fendley, Phys. Rev. B {\bf 66}, 144416
	(2002).


\bibitem{MS-nonabel}M. Sato, Phys. Rev. B {\bf 71}, 024402 (2005). 













\end{thebibliography}
\end{document}